\documentclass[12pt]{article}
\usepackage{amsmath}
\usepackage{amsthm}
\usepackage{amsfonts}
\usepackage{graphicx}
\usepackage{enumerate}
\usepackage[round, authoryear]{natbib}
\usepackage{url} 

\newcommand{\blind}{0}

\usepackage{bbm}
\usepackage[utf8]{inputenc}
\usepackage{subfigure} 
\usepackage{multirow} 
\usepackage{multicol} 
\usepackage{arydshln}
\usepackage{mathrsfs}

\usepackage{bm}
\newcommand{\Real}{\mathbb{R}}

\newcommand{\V}[1]{{\pmb{\bm{\mathbf{\MakeLowercase{#1}}}}}} 
\newcommand{\M}[1]{{\pmb{\bm{\mathbf{\MakeUppercase{#1}}}}}} 

\newcommand{\comment}[1]{}
\newcommand{\code}[1]{\texttt{#1}}
\def\code#1{\texttt{#1}}

\begin{document}

\def\spacingset#1{\renewcommand{\baselinestretch}%
{#1}\small\normalsize} \spacingset{1}


\if1\blind
{
  \title{\bf Length Penalty derivation}
  \author{Charlie 1\thanks{
    The authors gratefully acknowledge \textit{please remember to list all relevant funding sources in the unblinded version}}\hspace{.2cm}\\
    Department of Statistics, NCSU\\
    and \\
    Author 2 \\
    Department of ZZZ, University of WWW}
  \maketitle
} \fi

\if0\blind
{
  \bigskip
  \bigskip
  \bigskip
  \begin{center}
    {\LARGE\bf Quantile sheet estimator with shape constraints}
\end{center}
  \medskip
} \fi

\bigskip
\begin{center}
    \Large
        
    \vspace{0.4cm}
    \textbf{Charlie Song}

\end{center}
\bigskip
\begin{abstract}
 A quantile sheet is a global estimator for multiple quantile curves. A quantile sheet estimator is proposed to maintain the non-crossing properties for different quantiles. The proposed estimator utilizes SCOP: shape-constrained P-spline to enforce the non-crossing properties directly in construction. A local GCV parameter tunning algorithm is used for fast estimation results. Data simulation shows the proposed method and existing competitors can recover the underlying quantiles with comparable mean square error.
\end{abstract}

\noindent%
\vfill

\newpage
\spacingset{1.5} 
\section{Motivation}
Quantile regression(QR) or conditional quantile regression is an alternative to conditional mean regression introduced by \cite{koenker1978regression}. The reason to use conditional quantile instead of mean is because QR is more robust to outliers and it directly gives the reference for selected quantiles which can be interpreted as the confidence region. \\
QR seeks to find the conditional quantile function $Q_\tau(x)$ that minimizes the criteria:
\begin{equation}
    {R_{\tau ,\lambda }}\left[ Q_\tau \right] = \sum\limits_{i = 1}^n {{\rho _\tau }\left\{ {{y_i} - Q_\tau\left( {{x_i}} \right)} \right\}}  + \lambda P(Q_\tau) 
\end{equation}
where 
\begin{equation}
    {\rho _\tau }\left( u \right) = u\left\{ {\tau  - I\left( {u < 0} \right)} \right\}
\end{equation}
is the check function proposed in
\cite{koenker1978regression}. There are various methods to estimate the $Q(x)_\tau$, involving $L_1$ or $L_2$ penalties, parametric or nonparametric estimators. And optimization techniques are usually applied to solve the minimization problem, like in \cite{koenker1994quantile} and \cite{takeuchi2006nonparametric}.\\
Non-crossing is a desired property identified in \cite{he1997quantile}. When $Q_{\tau_k}(x)$ is estimated for various $\tau_k$, the resulting reference curves may cross or overlap, which contradicts the underlying assumption for conditional quantiles. Thus, we want to impose a constraint that $Q_\tau(x)$ is monotone non-decreasing in $\tau$.\\
Currently, most methods obtain non-crossing by considering several selected $\{\tau_k: k = 1,\cdots,K\}$ (order in increasing order) , and constrain $Q_{\tau_k}(x)$ such that $Q_{\tau_k}(x) > Q_{\tau_{k-1}}(x)$. However, we notice two major flaws in this kind of method: first, the estimated $Q_{\tau_k}(x)$ is not unique and prone to be different shapes for various $\{\tau_k: k = 1,\cdots,K\}$; second, the non-crossing property only applies to the set $\{\tau_k: k = 1,\cdots,K\}$. In other word, $Q_{\tau_l}(x)$ may violate the non-crossing property if $\tau_l \notin \{\tau_k: k = 1,\cdots,K\}$.\\
Therefore, we consider a conditional quantile estuator $\hat{Q}(x,\tau)$ minimizing the corresponding criteria:
\begin{equation}
    {R_{\lambda }}\left[ Q \right] = \frac{1}{n}\sum\limits_{i = 1}^n \int_0^1 {{\rho _\tau }\left\{ {{y_i} - Q\left( {{\tau,x_i}} \right)} \right\}} d\tau + \mathcal{P} 
\end{equation}

\section{Methods}
We are interested in imposing the monotone increasing constraint in $Q(x, \tau)$, so our model is properly defined.
\subsection{B-splines}
\vskip 2MM
\noindent
Using the notation in \citep{xiao2019asymptotic}, we state here the \textbf{Carl de Boor's recursion formula\citep{de1978practical}}:\\
 The $[m]$ order B-splines on a sequence of knots $\underbar{t}=\{ 0= t_0<t_1< \cdots< t_{K_0+1}=1\}$, with  ${N}^{[m]}_{k}(x) = \tilde{N}^{[m]}_{k-m}(x), \ 1 \leq k \leq K = K_0 +m $: 
\begin{align*}
    \tilde{N}^{[1]}_{k}(x)
    & = 
    \begin{cases}
        1 &\text{if $t_k \leq x  < t_{k+1}$}\\
        0 &\text{otherwise}\\
    \end{cases} \\
    \tilde{N}^{[m]}_{k}(x) 
    &= \frac{x-t_k}{t_{k+m-1}-t_k}\tilde{N}^{[m-1]}_{k}(x) + \frac{t_{k+m}-x}{t_{k+m}-t_{k+1}}\tilde{N}^{[m-1]}_{k+1}(x)
\end{align*}
for $k = -(m-1), \cdots, K_0$.\\
\noindent
There are $(K_0+m)$ B-splines functions of order $[m]$. And requires that ${0}/{0}=0$.\\
For convenience, denotes ${N}^{[m]}_{k}(x)$ by $N_k(x)$ and write $\M N(x) = [N_1(x), \cdots, N_K(x)]^T \in \Real^k$.
\subsubsection{P-spline}
The concept of P-spline comes from \citep{eilers1996flexible}, which uses a differencing matrix to penalize the smoothness of the model. The use of P-splines also requires the knots sequence to be equally spaced so that $\underbar{t}=\{ 0= t_0<t_1< \cdots< t_{K_0+1}=1\}=\{0, h, 2h, 3h, K_0h, 1\}$. Aternatively, we can write $t_{k} = kh$, with $h = 1 / (K_0 + 1)$.
\subsection{SCOP-splines \cite{pya2015shape}}
The details of B-splines are given in \cite{de1978practical}. To accommodate the smoothness and fidelity issue, \cite{eilers1996flexible} propose a penalized version of B-splines, now known as the popular P-spline. To achieve the desired shape constraint on the estimated curves, \cite{pya2015shape} reparametrized the coefficients of P-splines and proposed the SCOP-splines.
\subsubsection{One-dimensional case}
Suppose that we want to construct a monotonically increasing smooth $Q(x)$ using a B-spline basis,
\begin{equation*}
    Q(x) = \sum_{j=1}^K \gamma_j N_j(x),
\end{equation*}
where $K$ is the number of basis function, the $N_j$ are the B-spline basis on interval $[a,b]$ with equally spaced knots, and $\gamma_j$ are the spline coefficients.\\
Observe that: \textbf{Sufficient conditions for $Q^\prime(x)\geq 0$ is that $\gamma_j \geq \gamma_{j-1} \forall j$}. One way is to re-parametrize $\V\gamma$, so that:
\begin{equation*}
    \V\gamma = \V\Sigma \tilde{\V\beta}
\end{equation*}
where $\V\beta=[\beta_1,\beta_2,\cdots,\beta_K]^T$ and $\tilde{\V\beta}=[\beta_1, \exp{\beta_2}, \cdots, \exp{\beta_K}]$, and $\Sigma_{ij}=0$ if $i<j$ and $\Sigma_{ij}=1$ if $i\geq j$.\\
At last, with $N_{ij}=N_j(x_i)$, we can represent $\M Q= [Q(x_1), \cdots, Q(x_n)]^T$ as
\begin{equation*}
    \M Q = \M N \M \Sigma \tilde{\V \beta}.
\end{equation*}
\textbf{Penalty}
Penalize on $\V \beta$ starting from $\beta_2$ is equivalent to a second-order P-spline penalty. Thus the criteria are
\begin{equation*}
    \parallel \V y -  \M N \M \Sigma \tilde{\V \beta}\parallel^2 + \lambda\parallel \M D \V \beta  \parallel^2
\end{equation*}
where $\M D$ is $(K-2)\times K$ matrix or first order difference matrix without the first row.

\subsubsection{Multi-dimensional SCOP-splines}
To be able to apply in higher dimensional and account for the correlative relation between covariates, tensor product spline basis is considered. For example, we have two covariates $x, \tau$ and want to fit a tensor product splines with the number of knots $K= K_\tau, K_1$ and order $m = m_\tau, m_1$ on each covariate. We impose the non-decreasing constraints on $\tau$:
\begin{equation*}
    Q(\tau, x) = \M N^T (\tau, x) \V \gamma
\end{equation*}
where $\M N = \M N^{[m_\tau]}(\tau) \otimes \M N^{[m_1]}(x) \in \Real^{K_\tau K_1}$, $\M N^{[m_\tau]}(\tau) \in \Real^{K_\tau}$ is the basis spline vector for $\tau$, and $\M N^{[m_1]}(x) \in \Real^{K_1}$ is basis spline vector for $x$.\\
\noindent
The constraint is guaranteed by parametrizing
\begin{equation*}
    \V \gamma = \M \Sigma \tilde{\V \beta} \in \Real^{K_\tau K_1},
\end{equation*} 
where the $K_\tau K_1 \times K_\tau K_1$ matrix $\M \Sigma =  \M \Sigma_\tau\otimes \M I_{K_1}$,\\
and the $K_\tau \times K_\tau$ matrix ${\M \Sigma_\tau}_{ij} = 1$ if $i\geq j$ and $0$ otherwise.
\begin{align*}
    \tilde{\V \beta} & = \text{vec} (\M \Theta ^T)\\
    & = [\beta_{11}, \cdots, \beta_{1K_1}, \exp(\beta_{21}), \cdots, \exp(\beta_{2K_1}), \cdots, \exp(\beta_{K_\tau 1}), \cdots, \exp(\beta_{K_\tau K_1} )]^T
\end{align*}
\vspace{1cm}
where $\M \Theta = \begin{pmatrix} 
\beta_{11}&  \cdots& \beta_{1K_1}\\
\exp(\beta_{21})&  \cdots& \exp(\beta_{2K_1})\\
\vdots & & \vdots\\
\exp(\beta_{K_\tau 1})&  \cdots& \exp(\beta_{K_\tau K_1})
\end{pmatrix}$\\
\noindent
\textbf{Penalty}: instead of penalize $\tilde{\V \beta}$ directly, \cite{pya2015shape} penalize on $\V \beta = (\beta_{11}, \cdots, \beta_{K_1K_2})^T$
\begin{equation*}
    \mathcal{P} = \lambda_\tau \parallel \M D_\tau \V \beta\parallel^2 + \lambda_{11} \parallel \M D_{11} \V \beta\parallel^2 + \lambda_{12} \parallel \M D_{12} \V \beta\parallel^2
\end{equation*}
where $\M D_\tau = \M F_{K_\tau} \otimes \M I_{K_1}$, $\M D_{11} = \M E_{K_\tau} \otimes \M \Delta_{K_1, 2}$, and $\M D_{12} =  (\M I_{K_\tau} -\M E_{K_\tau}) \otimes \M \Delta_{K_1,1} $ ,\\
$\M \Delta_{K, q} \in \Real ^{(K-q) \times K}$ denotes the $q^{th}$ order difference operator\\
$\M E_{K_\tau} = \begin{pmatrix} 1 & & & & 0 \\
& & \ddots & &\\
0& & & & 0\end{pmatrix} \in \Real^{K_\tau \times K_\tau}$\\
$\M F_{K_\tau} \in \Real^{(K_\tau-2) \times K_\tau}$ is $\M \Delta_{K_\tau,1}$ without the first row.\\
To simplify the notation a little bit, we combine these two penalty terms into one:
\begin{equation*}
    \mathcal{P} = \V \beta^T \M S_{\V \lambda} \V \beta 
\end{equation*}
where $\M S_{\V \lambda} = \lambda_\tau \M D_\tau^T \M D_\tau + \lambda_{11} \M D_{11}^T \M D_{11} + \lambda_{12} \M D_{12}^T \M D_{12}$.\\
Usually set $\lambda_{11} = \lambda_{12}$ as $\M D_{11}, \M D_{12}$ both penalize the covariate $x$.
\subsection{Optimization criteria}
We focus on the optimization criteria:
\begin{align*}
    {R_{\lambda }}\left[ Q \right] 
    &=\mathcal{L}  + \mathcal{P} 
\end{align*}
where $\mathcal{L}$ is the loss of the quantile functions across the continuous domain of $\tau$. 
\begin{align*}
    \mathcal{L} 
    &= \frac{1}{n}\sum\limits_{i = 1}^n \int_0^1 {{\rho _\tau }\left\{ {{y_i} - Q\left( {{\tau,x_i}} \right)} \right\}} d\tau\\
    &= \frac{1}{n}\sum\limits_{i = 1}^n \int_0^1 \left( {{y_i} - Q\left( {{\tau,x_i}} \right)} \right) \cdot \left[\tau - I\left( {{y_i} < Q\left( {{\tau,x_i}} \right)} \right)\right] d\tau
\end{align*} 

\subsubsection{Calculate the gradient of $\mathcal{L}$}
Since the loss function $\mathcal{L}$ of quantile regression is not differentiable at $\V \beta$ when $y_i = Q(\tau,x_i)$, we approximate the gradient by holding the quantile function $Q(\tau,x_i)$ in the indicator function $I( y_i < Q(\tau,x_i) )$ fixed.\\
Let $\M C \in \Real^{K_\tau K_1 \times K_\tau K_1}$ a diagonal matrix depends on $\V \beta$, s.t.  
$
\M C_{jj} = \begin{cases}
        1, &\text{if } \tilde{\beta}_j=\beta_j\\
        \exp(\beta_j), &\text{otherwise}.
    \end{cases}
$
\begin{align*}
    \nabla \mathcal{L} (\V \beta) 
    & = - \frac{1}{n}\sum_{i=1} ^ n \int_0^1 \left[\tau - I\left( {{y_i} - Q\left( \tau,x_i \right)} \right)\right] \M C \M \Sigma^T \M N(\tau, x_i) d\tau\\
    & = - \frac{1}{n}\sum_{i=1} ^ n \int_0^1 \tau \M C \M \Sigma^T  \M N(\tau, x_i) d\tau - \int_0^1  I\left( {{y_i} < Q\left( \tau,x_i \right)} \right) \M C \M \Sigma^T \M N(\tau, x_i) d\tau\\
    & = - \M C \M \Sigma^T \frac{1}{n}\sum_{i=1} ^ n \int_0^1 \tau   \M N(\tau, x_i) d\tau - \int_0^1  I\left( {{y_i} < Q\left( \tau,x_i \right)} \right)  \M N(\tau, x_i) d\tau\\
    & = - \M C \M \Sigma^T\frac{1}{n}\sum_{i=1} ^ n \{A + B\}
\end{align*}
For equation $A$, using integration by parts in \citep{vermeulen1992integrating}
\begin{align*}
    A &=  \int_0^1 \tau   \M N(\tau, x_i) d\tau\\
    &=   \int_0^1 \tau   \M N(\tau, x_i) d\tau\\
    & =  \int_0^1 \tau  \M N^{[m_\tau]}(\tau) \otimes \M N^{[m_1]}(x_i) d\tau\\
    & =   \int_0^1 \tau  \M N^{[m_\tau]}(\tau) d\tau \otimes \M N^{[m_1]}(x_i) \\
    & =  \{\tau  \M G_1 \M \Sigma_\tau^T \M N^{[m_\tau+1]}(\tau) |_0^1 - \int_0^1 \M G_1 \M \Sigma_\tau^T \M N^{[m_\tau+1]}(\tau)  d\tau\} \otimes \M N^{[m_1]}(x_i)\\
    & = \{ \M G_1 \M \Sigma_\tau^T [\M N^{[m_\tau + 1]}(1)  -  \M G_2 \M \Sigma_\tau^T \M N^{[m_\tau+2]}(\tau)|_0^1 ]   \} \otimes \M N^{[m_1]}(x_i)
\end{align*}
where $\M N^{[m_\tau + 1]}(\tau), \M N^{[m_\tau + 2]}(\tau) \in \Real^{K_\tau}$ represent the B-splines basis vector of order $m=m_\tau+1, m_\tau+2$ constructed without the first one and two elements. The integration of B-splines is well known to be the B-splines function with one and two orders higher subject to some coefficients. Specifically, diagonal matrix $\M G_1, \M G_2 \in \Real^{K_\tau\times K_\tau}$ is defined as ${\M G_1}_{ii} = (t_{i+m_2} - t_{i})/m_\tau$ and ${\M G_2}_{ii} = (t_{i+m_2+1} - t_{i})/(m_\tau+1)$. For more detail see \citep{de1978practical}.\\

For equation $B$,
\begin{align*}
    B &= - \int_0^1  I\left( {{y_i} < Q\left( {{\tau,x_i}} \right)} \right)  \M N(\tau, x_i) d\tau\\
    & = -  \int_0^1  I\left( {{y_i} < Q\left( {{\tau,x_i}} \right)} \right)  \M N^{[m_\tau]}(\tau) \otimes\M N^{[m_1]}(x_i) d\tau\\
    & = -   \int_0^1  I\left( {{y_i} < Q\left( {{\tau,x_i}} \right)} \right)  \M N^{[m_\tau]}(\tau) d\tau \otimes \M N^{[m_1]}(x_i)\\
    & = -   \int_{\tau_i^*}^1   \M N^{[m_\tau]}(\tau) d\tau \otimes \M N^{[m_1]}(x_i)\\
    & = -    \M G_1 \M \Sigma_\tau^T \M N^{[m_\tau+1]}(\tau)|_{\tau^*}^1  \otimes \M N^{[m_1]}(x_i)
\end{align*}
where $Q( \tau_i^*, x_i ) = y_i$. $\tau_i^*$ depends on $\V \beta$, which could be interpreted as the estimated conditional cumulative probability given the quantile function $Q(  \tau,x )$ and $x_i$.\\
Therefore, the derivative can be written as:
\begin{align*}
    \nabla \mathcal{L} (\V \beta) 
    &= - \M C \M \Sigma^T\frac{1}{n}\sum_{i=1} ^ n  \{ \M G_1 \M \Sigma_\tau^T [ -  \M G_2\M \Sigma_\tau^T\M N^{[m_\tau+2]}(\tau)|_0^1 + \M N^{[m_\tau+1]}(\tau_i^*)]\} \otimes \M N^{[m_1]}(x_i)\\
    &= - \M C \M \Sigma^T\frac{1}{n}\sum_{i=1} ^ n \{ -\M G_1 \M \Sigma_\tau^T  \M G_2\M \Sigma_\tau^T\M N^{[m_\tau+2]}(\tau)|_0^1\}\otimes \M N^{[m_1]}(x_i)  +  \{ \M G_1 \M \Sigma_\tau^T\M N^{[m_\tau+1]}(\tau_i^*)]\} \otimes \M N^{[m_1]}(x_i)\\
    &= - \frac{1}{n}\M C \M \Sigma^T \{ [-\M G_1 \M \Sigma_\tau^T  \M G_2\M \Sigma_\tau^T\M N^{[m_\tau+2]}(\tau)|_0^1] \otimes\M N_{1}^T\V 1^{n\times 1} + [ \M G_1 \M \Sigma_\tau^T\M N_{\tau^*}^T]\V  \otimes_{col} \M N_{1}^T 1^{n\times 1}\}\\
    &= - \frac{1}{n}\M C \M \Sigma^T \{ \M H_1 + \M H_\tau\}
\end{align*}
where $\M N_1 = [\M N^{[m_1]}(x_1), \cdots, \M N^{[m_1]}(x_n)]^T \in \Real^{n\times K_1}$, $\M N_{\tau^*} = [\M N^{[m_\tau+1]}(\tau_1^*), \cdots, \M N^{[m_\tau+1]}(\tau_n^*)]^T \in \Real^{n\times K_\tau}$, $\otimes_{col}$ represent the column-wise Kronecker product operator.\\
$\M H_1 =  [-\M G_1 \M \Sigma_\tau^T  \M G_2\M \Sigma_\tau^T\M N^{[m_2+2]}(\tau)|_0^1] \otimes\M N_{1}^T\V 1^{n\times 1}$, $\M H_2 = [ \M G_1 \M \Sigma_\tau^T\M N_{\tau^*}^T]\V \otimes_{col} \M N_{1}^T 1^{n\times 1}.$
\subsubsection{Gradient of $R[Q]$}
Combining with the gradient of penalty, we obtain an approximation of the gradient of the optimization criterion $R[Q]$.
\begin{equation}
    \nabla R(\V \beta) = - \frac{1}{n}\M C \M \Sigma^T ( \M H_1 + \M H_2) + \M S_{\V \lambda} \V \beta
    \label{equ:deriv}
\end{equation}
\subsubsection{Hessian of $R[Q]$}
If we adopt the same approach by holding the quantile function $Q(x_i,\tau)$ in the indicator function $I\left( {{y_i} < Q\left( {{x_i,\tau}} \right)} \right)$ fixed, we can approximate the hessian matrix of $R[Q]$:
\begin{equation}
    \M H(\tilde{\V \beta})  = \frac{1}{n}\M J(\tilde{\V \beta}) + \M S_{\V \lambda}
    \label{equ:hess}
\end{equation}
where $\M J_{jj} = \begin{cases}
        0, &\text{if } \tilde{\beta}_j=\beta_j\\
        [- \M C \M \Sigma^T ( \M H_1 + \M H_2)]_j, &\text{otherwise}.
    \end{cases}$
However, the estimated hessian matrix is singular with a very large condition number. My experiments with Newton's method have not been successful if I use a generalized inverse in place of the inverse of the hessian matrix. In \citep{pya2015shape}, the authors discuss an approach to estimate its inverse by augmenting the hessian matrix and performing QR decomposition. Their idea is worth a try, but requires further implementations.
\subsubsection{Rewrite the criteria}
Using the notation above, we can rewrite the optimization criteria 
${R_{\lambda }}\left[ Q \right]$ as
\begin{equation}
    R(\V \beta) =  \frac{1}{n}(\V \tau^* - 0.5)^T \V y -  \frac{1}{n}( \M H_1 + \M H_2)^T \M \Sigma \tilde{\M \beta} + {\V \beta}^T \M S_{\V \lambda} {\V \beta}
 \label{equ:loss}
\end{equation}
where $\V \tau^* = 
\begin{pmatrix}
    \tau_1^* & \cdots & \tau_n^* 
\end{pmatrix}$
and $\V y = \begin{pmatrix}
    y_1 & \cdots & y_n 
\end{pmatrix}$
\subsection{Algorithm: gradient based}
The value $\M H_1$ is independent of $\V \beta$. Thus, we only need to calculate $\M H_1$ once and store it, then we can reuse it to calculate the next derivative. However, $\M C$ and $\M H_2$ does depends on $\V \beta$. We update their values iteratively.
\subsubsection{Initialization}
To obtain an initial estimate of $\V \beta$, we seek to minimize a penalized constrained least square problem:
\begin{equation}
    ls(\V \beta) = \parallel \V y - \M N (\tilde{\V \tau}, \V x) \M\Sigma \tilde{\V \beta} \parallel_2^2 + n\tilde{\V \beta}^T \M S_{\V \lambda} \tilde{\V \beta}
    \label{equ:leastsq}
\end{equation}
subject to linear inequality constraints that $\tilde{\beta}_j > 0$ whenever $\tilde{\beta}_j = \exp(\beta_j)$\\
The vector $\tilde{\V \tau}$ is a vector of estimated conditional probability for each observation using the local kernel method with a span set to be $0.1$ of the range of $x$.

\subsubsection{Stopping criteria}
Generally, the stopping criteria for the descent-based algorithm are set to be $\parallel \nabla R(\V \beta) \parallel_2 \leq \eta$. I found it might be more appropriate as the fraction of descents size out of the size of the coefficient, so I use $$\parallel \nabla R(\V \beta) \parallel_2/\parallel\V \beta \parallel_2 \leq \eta$$
where $\eta$ is small and positive.\\
Alternatively, we can use the decrease in loss function as the stopping criterion. 
\begin{equation*}
    \frac{|R(\V \beta^{[k]}) - R(\V \beta^{[k+1]})|}{|R(\V \beta^{[k]})|} \leq \epsilon
\end{equation*}
\subsubsection{Gradient descent: backtracking line search} \citep{boyd2004convex}
\vskip 2mm
\fbox{%
    \parbox{\textwidth}{%
        \textbf{Require:}  $\M N_1, \M G_1, \M G_2, \M \Sigma_\tau, \M S, \alpha \in (0, 0.5), \beta \in (0,1).$ \\
        Precalculate $\M H_1$,\\
        \textbf{Initialize} a starting point for $\V \beta$\\
        \textbf{repeat}\\
        $1\quad\quad$ $\Delta \V \beta = -\nabla R(\V \beta)$ using \eqref{equ:deriv}\\
        $2\quad\quad$ Line search. Choose step size $t$ via backtracking line search:\\
        $:\quad\quad\quad$ $t:=1.$ \\
        $:\quad\quad\quad$ \text{While} $f(\V \beta+t\Delta \V \beta) > f(\V \beta) + \alpha t \nabla f(\V \beta)^T \Delta \V \beta,$ $\quad t:=\beta t.$ \\
        $3\quad\quad$ Update $\V \beta := \V \beta + t \Delta \V \beta$.\\
        \textbf{Until} stopping criterion is satisfied.
    }%

}

According to \citep{boyd2004convex}, the parameter $\alpha$ is typically within $(0.01, 0.3)$ meaning we accept a decrease in loss function between $1\%$ and $30\%$ of prediction based on linear extrapolation. The parameter $\beta$ is often within $(0.1, 0.8)$, which corresponds from a crude search to a finer search.
\subsubsection{Experiment}
\begin{itemize}
    \item \textbf{Tunning:} From my experiment, I found it hard to set the appropriate $\alpha$ and $\beta$ to get the algorithm to converge, because it spends too much time searching the right $t$(The convergence is guaranteed by forcing the step size $t$ to be no less than a specified value). The Barzalai-borwein stepsize is not converging for the setup, possibly because the QR loss function is not twice differentiable.
    \item \textbf{Time:} The most time consuming steps comes from calculating the $\V \tau^*$ at each iteration in order to calculate $\M N_{\tau^*}$ and $\M H_2$. Current implementation loops through each $x_i$ to get the corresponding $\tau^*_i$; I expect a major time decrease once I implement it on \code{Fortran}.
\end{itemize}

\subsection{Smoothing Quantile Regression}
The smoothing Quantile Regression framework proposed by \citep{fernandes2021smoothing} gives us a new perspective toward quantile regression. This new framework resolves the non-differentiability problem of the check function of the classical QR loss functions and provides a twice differentiable and locally strong convex loss function, which facilitates a faster convergence rate and lower estimation error. \citep{he2020smoothed} conducted an extensive study of the proposed smoothing QR framework on large dimension regime and points out that the new method allows Quasi-Newton gradient-based optimization and proposed gradient methods with Barzalai-Borwein\citep{barzilai1988two} step size. In comparison with the classical QR, the new framework has an estimator and inference method that is not worse in estimation accuracy and far better scalability when the dimension is large.
In light of the smoothed quantile regression framework, from an $M$-estimation point of view, we can write our new loss function as
\begin{equation}
    \mathcal{L}_{h} = \frac{1}{n}\sum\lim_{i = 1}^n \int_0^1 {{\mathbb{l}_{h,\tau} }\left\{ {{y_i} - Q\left( {{\tau,x_i}} \right)} \right\}} d\tau
\end{equation}
with $\mathbb{l}_{h,\tau}(u) = (\rho_\tau * K_h)(u) = \int_{-\infty}^{\infty} \rho_\tau(v)K_h(v-u) dv,$\\
$K_h(u) = h^{-1}K(u/h)$ and
$K(\cdot)$ is a kernel function integrate to 1, and $h>0$ is a bandwidth value.\\
This corresponding estimator is also referred to as  $conquer$ in \citep{he2020smoothed}.

\subsection{Gradient and Hessian for smoothed QR}
The convolution-type kernel smoothing loss function is twice continuously differentiable. The gradient vector can be written using the notation above:
\begin{align*}
    \nabla \mathcal{L}_h 
    & = \frac{1}{n}\M C \M \Sigma^T \sum_{i=1}^n \int^1_0 \{\mathcal{K}_h[Q(\tau,x_i)-y_i]-\tau\} \M N(\tau, x_i) d\tau
\end{align*}
where $\mathcal{K}_h(u) = \int_{-\infty}^{u/h} K(v) dv$. We can write it together:
\begin{align}
    \nabla \mathcal{L}_h 
    & = \frac{1}{n}\M C \M \Sigma^T \sum_{i=1}^n \int^1_0 \{\mathcal{K}_h[Q(\tau,x_i)-y_i]-\tau\} \M N(\tau, x_i) d\tau\\
    & = \frac{1}{n}\M C \M \Sigma^T \sum_{i=1}^n \int^1_0 \{\int_{-\infty}^{[Q(\tau,x_i)-y_i]/h} K(v) dv-\tau\} \M N(\tau, x_i) d\tau\\
    & = \frac{1}{n}\M C \M \Sigma^T \sum_{i=1}^n \int^1_0 \int_{-\infty}^{[Q(\tau,x_i)-y_i]/h} K(v)\M N(\tau, x_i) dv  d\tau-\int^1_0 \tau \M N(\tau, x_i) d\tau\\
    & = \frac{1}{n}\M C \M \Sigma^T \sum_{i=1}^n \int^1_0 \int_{-\infty}^{[\M N^T (\tau, x_i) \M \Sigma \tilde{\V \beta}-y_i]/h} K(v)\M N(\tau, x_i) dv  d\tau-\int^1_0 \tau \M N(\tau, x_i) d\tau\\
    & = \frac{1}{n}\M C \M \Sigma^T [\V h_\tau - \V h_1]
    \label{equ:sgrad}
\end{align}
if we write $\V h_\tau = \sum_{i=1}^n \int^1_0 \int_{-\infty}^{[Q(\tau,x_i)-y_i]/h} K(v) \M N(\tau, x_i)dv d\tau$,\\
and $\V h_1 = \sum_{i=1}^n \int^1_0 \tau \M N(\tau, x_i) d\tau$.\\
$\V h_1$ is the same as before, but $\V h_\tau$ cannot be computed analytically because the integrand is a compound function of $\tau$. We proceed to estimate $\V h_\tau$ by numerical integration, that is to evaluate $\int_{-\infty}^{[Q(\tau,x_i)-y_i]/h} K(v) \M N(\tau, x_i)dv$ at $n_\tau$ equal-spaced value in $[0,1]$ and calculate their mean.\\
The hessian matrix for smoother QR is:
\begin{align*}
    \nabla^2 \mathcal{L}_h 
    & = \M C \M \Sigma^T \M W \M \Sigma \M C + \M J\\
\end{align*}
where $\M W = \frac{1}{n}\sum_{i=1}^n \int^1_0 K_h[Q(\tau,x_i)-y_i] \M N(\tau, x_i)\M N^T(\tau, x_i) d\tau$,\\
and the diagonal matrix 
$\M J_{jj} = \begin{cases}
        0, &\text{if } \tilde{\beta}_j=\beta_j\\
        [\frac{1}{n}\M C \M \Sigma^T [\V h_\tau - \V h_1]]_j, &\text{otherwise}.
    \end{cases}$
    
\subsection{Barzilai-Borwein Stepsize}
Since we know the hessian matrix for coefficients, we can employ Newton updates algorithms to minimize the objective functions. However, the calculation of matrix $\M W$ is computationally intensive and requires high-order numerical integration; and the inversion of the hessian matrix for every iteration is also expensive. Therefore, we prefer a first-order update scheme, like a gradient-based method. We are able to use the fact that a hessian exists and the objective function convex by using a Quasi-Newton method as in \citep{he2020smoothed}.\\
The Brazilai-Borwein step size calculation:
\begin{equation*}
    \eta_{1,t} = \frac{\langle\V \delta^t, \V \delta^t \rangle}{\langle \V \delta^t, \V g^t \rangle}, \quad
    \eta_{2,t} = \frac{\langle\V \delta^t, \V g^t \rangle}{\langle \V g^t, \V g^t \rangle}
\end{equation*}
where $\delta^t = \V\beta^t - \V\beta^{t-1}$, $g^t = \nabla R(\V \beta^t) - \nabla R(\V \beta^{t-1})$ for $t = 1,2,\cdots$\\
The step size is then chosen with an upper bound $u$ if $\eta_{1,t}  >0$, and 1 otherwise.:
\begin{equation*}
    \eta_t = \min\{\eta_{1,t}, \eta_{2,t}, u\}.
\end{equation*}
\textbf{Initialization:} We will need to initialize $\V \beta^0$, and $\V \beta^1$ is computed by standard gradient descent or backtracking line searched step size.\\
\textbf{Stopping:} Usually the algorithm stop when the estimated gradient at step $t$ is less than a threshold:
\begin{equation*}
    \parallel \nabla R(\V \beta^{t}) \parallel_2 < \delta
\end{equation*}
provided that $\delta \leq \sqrt{p/n}$

\subsection{Gradient descent: GD-BB} \citep{barzilai1988two}
\vskip 2mm
\fbox{%
    \parbox{\textwidth}{%
        \textbf{Require:}  $\V h_1, \M S,$ bandwidth $h\in(0,1)$, gradient tolerance $\delta$, maximum step size $u$.\\
        \textbf{Initialize} a starting point for $\V \beta^0$\\
        Compute $\V \beta^1 = \V \beta^0 - \eta_{0}\nabla R(\V \beta^0)$\\
        \textbf{for $t = 1,2,\cdots$ do}\\
        $1\quad\quad$ $\Delta \V \beta = -\nabla R(\V \beta)$ using \eqref{equ:sgrad}\\
        $2\quad\quad$ BB stepsize. Choose step size $t$ :\\
        $:\quad\quad\quad$ $\delta^t = \V\beta^t - \V\beta^{t-1}$, $g^t = \nabla R(\V \beta^t) - \nabla R(\V \beta^{t-1})$ \\
        $:\quad\quad\quad$ $\eta_{1,t} = \langle\V \delta^t, \V \delta^t \rangle/\langle \V \delta^t, \V g^t \rangle,
    \eta_{2,t} = \langle\V \delta^t, \V g^t \rangle/\langle \V g^t, \V g^t \rangle$. \\
        $:\quad\quad\quad$ $\eta_t \leftarrow \min\{\eta_{1,t}, \eta_{2,t}, u\}$ if $\eta_{1,t} > 0$ and $\eta_t \leftarrow 1$ otherwise.\\
        $3\quad\quad$ $\V \beta^{t+1} = \V \beta^t + \eta_t \Delta \V \beta$\\
        \textbf{end for} when stopping criterion is satisfied, i.e. $\parallel \nabla R(\V \beta^{t}) \parallel_2 < \delta$
    }%

}
\subsection{Quantile Sheets \citep{schnabel2013simultaneous}}
The idea in \citep{schnabel2013simultaneous} is highly similar to our research methods, both of us consider the conditional probability $\tau$ as a covariate in regression function $Q(\tau, x)$; both of us use tensor product spline to estimate the quantile curves across various $\tau \in [0, 1]$. However, our methods differ in (1) we considered a constrained version of the tensor product spline which guarantees that different quantile curves will not cross, (2) we seek to minimize the $L_1$ regression directly while the authors use Schlossmacher's iterative reweight least square algorithm (IRLS) \cite{schlossmacher1973iterative}, (3) we treat $\tau$ as smooth as possible, so we integrate the objective function in $\tau$ analytically. This is for the consideration of both numerical and estimation efficiency. \citep{schnabel2013simultaneous} on the other hand, select a few $\tau$s and numerically integrate the objective function, their method is computationally inefficient and could be seen as a weighted version of analytical integration.\\
We combine the modified schlossmacher's IRLS algorithm with constrained tensor product spline using the package `\code{scam}' to create the so-called \textbf{constrained quantile sheet} (CQS).
\subsection{Two-step Ad-hoc constrained quantile regression}
The last method we consider in this monograph is the two-step ad-hoc constrained quantile regression. It is natural to estimate the conditional quantile at each $x$ first, and then use a least square method to regress the quantile curves. This idea, however, is subject to a huge challenge: (1) Using the least square will be vulnerable to outliers and damage the robustness of $L_1$ regression, (2) the estimated conditional quantile directly influences the final outcome of the quantile curves, but there is no standard procedure for estimating conditional quantile, and different estimation methods involve various parameters, (3) this estimator does not carry the so-called quantile properties, and there is no known theory guaranteeing the process is unbias.
\section{Simulation}
We run simulation studies to access the performance of our proposed methods compared to existing methods. The goal of the simulation studies is three folds: first, we want to compare how well our proposed method recovers the underlying quantiles for $\tau \in [0,1]$ by comparing the mean square errors; second, we want to investigate how well each method deal with the quantile crossing issue by counting the crossing occurrence; third, we are interested in the effect of a penalty and propose a way to tune the smoothing parameters.\\
The simulation data are generated according to the model 
\begin{equation*}
    y_i = g(x_i) + \sigma(x_i) \epsilon_i,
\end{equation*}
where covariate $x_i$ is generated from uniform distribution $U(0,1)$. We genenrate the signal $g(x_i)$ is 5 different schemes: (1) linear $g_1 = 0.2 + 0.4 x_i$, (2) logarithm $g_2 = \log(x_i)$, (3) sinusoidal $g_3 = \sin{(2\pi x_i)}$, (4) `linear sinusoidal' $g_4 = 0.5 + 2x_i + \sin{(2\pi x_i -0.5)}$ and (5) `square root sinusoidal' $g_5 = \sqrt{x_i(1-x_i)} \sin([2\pi(1+2^{-7/5}]/(x_i + 2^{-7/5}))$. The random noise $\epsilon_i$ is generated from 5 distributions: (i) Gaussian distribution $\mathcal{N}(0,1)$, (ii) t distribution with 3 degrees of freedom $t_3$, (iii) t distribution with 1 degree of freedom $t_1$, (iv) double exponential or Laplace distribution and (v) chi-square distribution with 3 degrees of freedom $\chi^2_3$. We consider three types of scale function, including homogeneous and heterogeneous models: (a) Constant (homogeneous model) $\sigma(x_i) = 0.2$, (b) Linear heterogeneous model $\sigma(x_i) = 0.2 (1 + x_i)$, (c) Quadratic heterogeneous model $\sigma(x_i) = 0.5 [1  + (x_i - 1 )^2 ]$. \\
The above simulation setting is modified from \citep{muggeo2013estimating} \citep{muggeo2020multiple}, \citep{fernandes2021smoothing}, \citep{he2020smoothed}. Because we are comparing methods that could estimate multiple quantiles from a single data set, we do not recenter the error $\epsilon_i$ at a $\tau$ quantile. However, we may center the error at the median.\\
We consider sample size at $\{64, 128, 256, 512\}$ with 100 replications for each combination of scenarios. We compare the constrained quantile sheet method with 4 existing methods: (1) \citep{schnabel2013simultaneous} quantile sheets (QS), (2) \citep{koenker1994quantile} piece-wise linear nonparametric quantile estimator (QRSS) in package `\code{quantreg}' as a reference, (3) \citep{muggeo2013estimating}, \citep{muggeo2020multiple} the auto-tuned growth-charts quantile regression (GCRQ) in the package `\code{quantregGrowth}', and one Ad-hoc method: two-step constrained quantile regression (cqreg). The methods of direct-constrained quantile regression and smoothing-constrained quantile regression are omitted due to poor performance. At $1024$ equally spaced quantiles level $\tau_j \in [0,1]$, the mean integrated square error (MISE($\tau_j$)) is evaluated at $10000$ equi-distant $x$. The number of crossing for neighboring quantile lines and estimation time is also compared.
\if1\blind
{

\section{Conclusion}
\label{sec:conc}

\bigskip
\begin{center}
{\large\bf SUPPLEMENTARY MATERIAL}
\end{center}

\begin{description}

\item[Title:] Brief description. (file type)

\item[R-package :] 

\item[HIV data set:] 

\end{description}

} \fi

\section{Reference}
\bibliographystyle{agsm}

\bibliography{Reference}

@article{eilers1996flexible,
  title={Flexible smoothing with B-splines and penalties},
  author={Eilers, Paul HC and Marx, Brian D and others},
  journal={Statistical science},
  volume={11},
  number={2},
  pages={89--121},
  year={1996},
  publisher={Institute of Mathematical Statistics}
}

@book{de1978practical,
  title={A practical guide to splines},
  author={De Boor, Carl},
  volume={27},
  year={1978},
  publisher={springer-verlag New York}
}

@article{xiao2019asymptotic,
  title={Asymptotic theory of penalized splines},
  author={Xiao, Luo and others},
  journal={Electronic Journal of Statistics},
  volume={13},
  number={1},
  pages={747--794},
  year={2019},
  publisher={The Institute of Mathematical Statistics and the Bernoulli Society}
}

@article{pya2015shape,
  title={Shape constrained additive models},
  author={Pya, Natalya and Wood, Simon N},
  journal={Statistics and Computing},
  volume={25},
  number={3},
  pages={543--559},
  year={2015},
  publisher={Springer}
}

@article{muggeo2013estimating,
  title={Estimating growth charts via nonparametric quantile regression: a practical framework with application in ecology},
  author={Muggeo, Vito MR and Sciandra, Mariangela and Tomasello, Agostino and Calvo, Sebastiano},
  journal={Environmental and ecological statistics},
  volume={20},
  number={4},
  pages={519--531},
  year={2013},
  publisher={Springer}
}

@article{muggeo2020multiple,
  title={Multiple smoothing parameters selection in additive regression quantiles},
  author={Muggeo, Vito MR and Torretta, Federico and Eilers, Paul HC and Sciandra, Mariangela and Attanasio, Massimo},
  journal={Statistical Modelling},
  pages={1471082X20929802},
  year={2020},
  publisher={SAGE Publications Sage India: New Delhi, India}
}

@article{koenker1994quantile,
  title={Quantile smoothing splines},
  author={Koenker, Roger and Ng, Pin and Portnoy, Stephen},
  journal={Biometrika},
  volume={81},
  number={4},
  pages={673--680},
  year={1994},
  publisher={Oxford University Press}
}

@article{koenker1978regression,
  title={Regression quantiles},
  author={Koenker, Roger and Bassett Jr, Gilbert},
  journal={Econometrica: journal of the Econometric Society},
  pages={33--50},
  year={1978},
  publisher={JSTOR}
}

@article{he1997quantile,
  title={Quantile curves without crossing},
  author={He, Xuming},
  journal={The American Statistician},
  volume={51},
  number={2},
  pages={186--192},
  year={1997},
  publisher={Taylor \& Francis}
}

@article{takeuchi2006nonparametric,
  title={Nonparametric quantile estimation},
  author={Takeuchi, Ichiro and Le, Quoc and Sears, Timothy and Smola, Alexander and others},
  year={2006},
  publisher={MIT Press}
}

@article{vermeulen1992integrating,
  title={Integrating products of B-splines},
  author={Vermeulen, Alan H and Bartels, Richard H and Heppler, Glenn R},
  journal={SIAM journal on scientific and statistical computing},
  volume={13},
  number={4},
  pages={1025--1038},
  year={1992},
  publisher={SIAM}
}

@book{boyd2004convex,
  title={Convex optimization},
  author={Boyd, Stephen and Boyd, Stephen P and Vandenberghe, Lieven},
  year={2004},
  publisher={Cambridge university press}
}

@article{fernandes2021smoothing,
  title={Smoothing quantile regressions},
  author={Fernandes, Marcelo and Guerre, Emmanuel and Horta, Eduardo},
  journal={Journal of Business \& Economic Statistics},
  volume={39},
  number={1},
  pages={338--357},
  year={2021},
  publisher={Taylor \& Francis}
}

@article{he2020smoothed,
  title={Smoothed quantile regression with large-scale inference},
  author={He, Xuming and Pan, Xiaoou and Tan, Kean Ming and Zhou, Wen-Xin},
  journal={arXiv preprint arXiv:2012.05187},
  year={2020}
}

@article{barzilai1988two,
  title={Two-point step size gradient methods},
  author={Barzilai, Jonathan and Borwein, Jonathan M},
  journal={IMA journal of numerical analysis},
  volume={8},
  number={1},
  pages={141--148},
  year={1988},
  publisher={Oxford University Press}
}

@article{schnabel2013simultaneous,
  title={Simultaneous estimation of quantile curves using quantile sheets},
  author={Schnabel, Sabine K and Eilers, Paul HC},
  journal={AStA Advances in Statistical Analysis},
  volume={97},
  number={1},
  pages={77--87},
  year={2013},
  publisher={Springer}
}

@article{schlossmacher1973iterative,
  title={An iterative technique for absolute deviations curve fitting},
  author={Schlossmacher, EJ},
  journal={Journal of the American Statistical Association},
  volume={68},
  number={344},
  pages={857--859},
  year={1973},
  publisher={Taylor \& Francis}
}
\end{document}